# Self-referenced characterization of optical frequency combs and arbitrary waveforms using a simple, linear, zero-delay implementation of spectral shearing interferometry

V.R. Supradeepa*, Christopher M. Long, Daniel E. Leaird and Andrew M. Weiner

*School of Electrical and Computer Engineering, Purdue University, West Lafayette, Indiana 47907, USA.*

*Corresponding author: supradeepa@purdue.edu*

**Abstract**: We discuss a simple, linear, zero-delay implementation of spectral shearing interferometry for amplitude and phase characterization of optical frequency comb sources and arbitrary waveforms. We demonstrate this technique by characterizing two different high repetition rate (~10 GHz) frequency comb sources, generated respectively by strong external and intracavity phase modulation of a continuous-wave laser. This technique is easy to implement, requiring only an intensity modulator and an optical spectrum analyzer (OSA), and is demonstrated to work at average power levels down to 100nW (10aJ/pulse at 10 GHz). By exploiting the long coherence lengths of these frequency combs and the self-referenced nature of the measurement, we also demonstrate a simple single-ended measurement of dispersion and dispersion slope in long lengths of fiber (>25km).

**References and links**

1. Z. Jiang, C.-B. Huang, D.E. Leaird, A.M. Weiner, "Optical arbitrary waveform processing of more than 100 spectral comb lines," Nature Photonics **1**, 463-467 (2007).
2. M. Kourogi, K. Nakagawa and M. Ohtsu "Wide-span optical frequency comb generator for accurate optical frequency difference measurement," IEEE J. Quantum Electron. **29**, 2693, (1993).
3. H. Murata, A. Morimoto, T. Kobayashi, and S. Yamamoto, "Optical pulse generation by electroopticmodulation method and its application to integrated ultrashort pulse generators," IEEE J. Sel. Top. QuantumElectron. **6**, 1325-1331 (2000).
4. P. Del'Haye, A. Schliesser, O. Arcizet, T. Wilken, R. Holzwarth and T. J. Kippenberg, "Optical frequency comb generation from a monolithic microresonator," Nature **450**, 1214-1217 (2007).
5. C.-B. Huang, S.-G. Park, D.E. Leaird, and A.M. Weiner, "Nonlinearly broadened phase-modulated continuous-wave laser frequency combs characterized using DPSK decoding," Opt. Express **16**, 2520-2527 (2008).
6. Seth M. Foreman, Kevin W. Holman, Darren D. Hudson, David J. Jones, and Jun Ye, "Remote transfer of ultrastable frequency references via fiber networks," Rev. Sci. Instrum. **78**, 021101 (2007).
7. F. Narbonneau, M. Lours, S. Bize, A. Clairon, G. Santarelli, O. Lopez, Ch. Daussy, A. Amy-Klein, and Ch. Chardonnet, "High resolution frequency standard dissemination via optical fiber metropolitan network," Rev. Sci. Instrum. **77**, 064701 (2006).
8. A.D. Ellis; F.C.G. Gunning, "Spectral density enhancement using coherent WDM," Photonics Technology Letters, IEEE **17**, 504-506 (2005).
9. R. Trebino, "Frequency resolved optical gating: The measurement of ultrashort optical pulses," Springer publications (2002).

10. R. P. Scott, N. K. Fontaine, Jing Cao, Katsu Okamoto, B. H. Kolner, J. P. Heritage, and S.. J. Ben Yoo, "High-fidelity line-by-line optical waveform generation and complete characterization using FROG," Opt. Express **15**, 9977-9988 (2007)
11. Ian A. Walmsley and Christophe Dorrer, "Characterization of ultrashort electromagnetic pulses," Adv. Opt. Photon. **1**, 308-437 (2009).
12. Christophe Dorrer and Inuk Kang, "Highly sensitive direct characterization of femtosecond pulses by electro-optic spectral shearing interferometry," Opt. Lett. **28**, 477-479 (2003)
13. V. R. Supradeepa, Daniel E. Leaird, and Andrew M. Weiner, "Optical arbitrary waveform characterization via dual-quadrature spectral interferometry," Opt. Express **17**, 25-33 (2009)
14. Nicolas K. Fontaine, Ryan P. Scott, Jonathan P. Heritage, and S. J. B. Yoo, "Near quantum-limited, single-shot coherent arbitrary optical waveform measurements," Opt. Express **17**, 12332-12344 (2009)
15. V. R. Supradeepa, Daniel E. Leaird, and Andrew M. Weiner, "Single shot amplitude and phase characterization of optical arbitrary waveforms," Opt. Express **17**, 14434-14443 (2009)
16. Jacob Cohen, Pamela Bowlan, Vikrant Chauhan, and Rick Trebino, "Measuring temporally complex ultrashort pulses using multiple-delay crossed-beam spectral interferometry," Opt. Express 18, 6583-6597 (2010)
17. Houxun Miao, Daniel E. Leaird, Carsten Langrock, Martin M. Fejer, and Andrew M. Weiner, "Optical arbitrary waveform characterization via dual-quadrature spectral shearing interferometry," Opt. Express **17**, 3381-3389 (2009)
18. Jean Debeau, Benoît Kowalski, and Rémi Boittin, "Simple method for the complete characterization of an optical pulse," Opt. Lett. **23**, 1784-1786 (1998).
19. M. Kwakernaak, R. Schreieck, A. Neiger and H. Jäckel, "Spectral phase measurement of mode-locked diode laser pulses by beating sidebands generated by electrooptical mixing", IEEE Photon. Technol. Lett., **12**, 1677-1679, (2000).
20. Z. Jiang, D. Leaird, C. B. Huang, H. Miao, M. Kourogi, K. Imai, and A. M. Weiner, "Spectral line-by-line pulse shaping on an optical frequency comb generator," IEEE J. Quantum Electron. **43**, 1163-1174 (2007).
21. V. R. Supradeepa, C. M. Long, D. E. Leaird, A. M. Weiner, "Fast Characterization of Dispersion and Dispersion Slope of Optical Fiber Links Using Spectral Interferometry With Frequency Combs," IEEE Photon. Technol. Lett **22**, 155-157 (2010).
22. http://www.aragonphotonics.com/ficha.php?cat=76&id=80&opt=2

## 1. Introduction

In recent years there has been significant work to characterize optical waveforms emanating from high repetition rate frequency combs. A key area driving the need for new measurement schemes is optical arbitrary waveform generation (OAWG) [1], wherein individual comb lines are controlled with arbitrary user defined phase and amplitude. When the pulse shaping function is not rapidly changing, i.e., static or quasi-static, OAWG generates periodic, very high complexity user defined waveforms (preserving the comb structure in the frequency domain) with applications ranging from communications and LIDAR to spectroscopy.

From a source based perspective, a main motivating factor has been recent activity towards development of techniques based on modulated continuous wave lasers to generate with relative ease, high repetition rate frequency combs which maintain optical frequency stability [2-5]. Although such "novel comb sources" may provide wide optical bandwidth, they do not generate short pulse outputs directly due to abrupt spectral phase variations. Understanding their spectral phase and amplitude characteristics is essential to know their time-frequency properties which in turn are necessary for their proper use and control. However, as we will discuss shortly, such combs pose unique challenges to characterization with conventional methods. Other areas which would benefit from new measurement capabilities for frequency comb waveforms include transmission of timing information over fiber links using frequency combs [6, 7] and coherent (carrier-phase-locked) WDM transmission formats where the relative phases between individual carriers become important, e.g. [8].

Conventional implementations of ultrafast measurement techniques, such as FROG [9, 10] and SPIDER and other spectral shearing interferometry variants [11, 12], although widely applied, have certain disadvantages for application to OAWG waveforms with extended time apertures. These include relatively long acquisition times due to delay scanning in FROG and very high spectral resolution requirements in SPIDER. Also, owing to the high repetition rate of these comb sources, the peak power for a given average power is low, and hence it is undesirable to rely on nonlinear optical effects for the measurement scheme. Recent work that addresses characterization of such waveforms includes adaptations of spectral interferometry [13-16], which is fast and linear but not self-referenced, and of spectral shearing interferometry [17], which is self-referenced and fast but requires nonlinear optical effects. In this work we will utilize an easy-to-use, zero-delay, and linear adaptation of spectral shearing interferometry to characterize the time-frequency behavior of two popular novel frequency comb sources, namely a comb generated by cascaded intensity and strong phase modulation of a CW laser [5] and one generated by an optical frequency comb generator (OFCG) [2]. Further, we show that by exploiting the long coherence length inherent to frequency combs, this characterization technique permits self-referenced, coherent dispersion and dispersion slope characterization of >25km lengths of optical fiber without requiring any external oscillator.

In conventional spectral shearing interferometry (SSI), phase information appears as amplitude modulation on the interference spectrum between the waveform and its spectrally sheared replica, which are separated by a delay greater than the waveform aperture. This delay is needed for unambiguous phase reconstruction. In the interference spectrum, there is an amplitude component caused by the phase difference between the waveforms which can be represented as the *cosine* or *sine* function of the phase difference. In such a case, there is a fundamental ambiguity in the inversion to obtain the actual phase difference, namely the two different solutions each of the above functions can give in the $(0,2\pi)$ range. However, if there exists a time delay term longer than the temporal length of the waveforms, the fringes which occur in the spectral interference pattern allow for overcoming this ambiguity. More details on this and the reconstruction procedure can be found in [11]. This however limits the maximum duty cycle of the waveforms to under 50% since any repetitive waveform with longer than 50% duty cycle will need time delays of more than half the period to separate them, but this just creates overlap with the next period of the waveform. Hence this makes it unusable for 100% duty cycle OAWG waveforms. A zero delay adaptation of SSI which avoids ambiguity in phase retrieval by obtaining the complex interference signal was introduced in [17] for OAWG applications. However this approach still requires nonlinearity to achieve spectral shearing. Here we discuss a different zero delay implementation of SSI in which spectral shearing is achieved in a linear fashion using an electro-optic intensity modulator. This allows for drastically simplifying the experimental setup. We utilize a modified adaptation of an elegant technique first proposed by J. Debeau et al in [18] for characterization of a gain-switched laser diode. Here we apply the technique to a variety of frequency comb and arbitrary waveforms and demonstrate excellent agreement with independent measurement techniques.

**2. Experimental setup**

Fig 1(a) shows the experimental setup. The input waveform is sent to a LiNOb3 Mach-Zehnder intensity modulator, which is driven by a weak sinusoidal RF signal at half the repetition frequency of the input frequency comb. For the data of Figs. 1-3, the half frequency drive signal was derived from a local oscillator which provides the RF signal for comb generation, while for the fiber experiments of Fig. 4 it was derived via clock recovery from the signal itself. The drive signal passes through a tunable RF phase shifter. Two spectra are recorded by the OSA with the RF phase shifter settings spaced 45° apart (with respect to the frequency halved signal). This could be implemented as a parallel measurement using a 0-45° RF hybrid, two intensity modulators, and a dual-channel spectrometer, e.g., [13]. From these two OSA spectra, both spectral amplitude and phase can be retrieved unambiguously. To understand how this technique works, let us first consider only a single optical frequency which goes through the intensity modulator. Assuming a small RF signal, only first order sidebands at half the repetition rate are generated on either side of the carrier. What the RF phase shift does is change the relative phase difference between first order sidebands as shown in fig 1(b). When the RF phase shifter moves by 45°, the phase difference between the sidebands moves by 90°. As shown in fig 1(c), with a frequency comb, at every possible sideband position contributions from two adjacent comb lines (at higher and lower optical frequency) will interfere (owing to the modulation frequency being exactly half the repetition rate), yielding sideband intensities that depend on the phase difference. As we discussed previously, since with only one interference spectrum we will have ambiguities w.r.t to retrieved phase, we use two spectra acquired sequentially at two states of the RF phase shifter. This provides both the in-phase and quadrature information of the interference, allowing for unambiguous phase retrieval. Simultaneous to acquiring the sideband spectra, the spectra at the carrier positions provide spectral amplitude information.

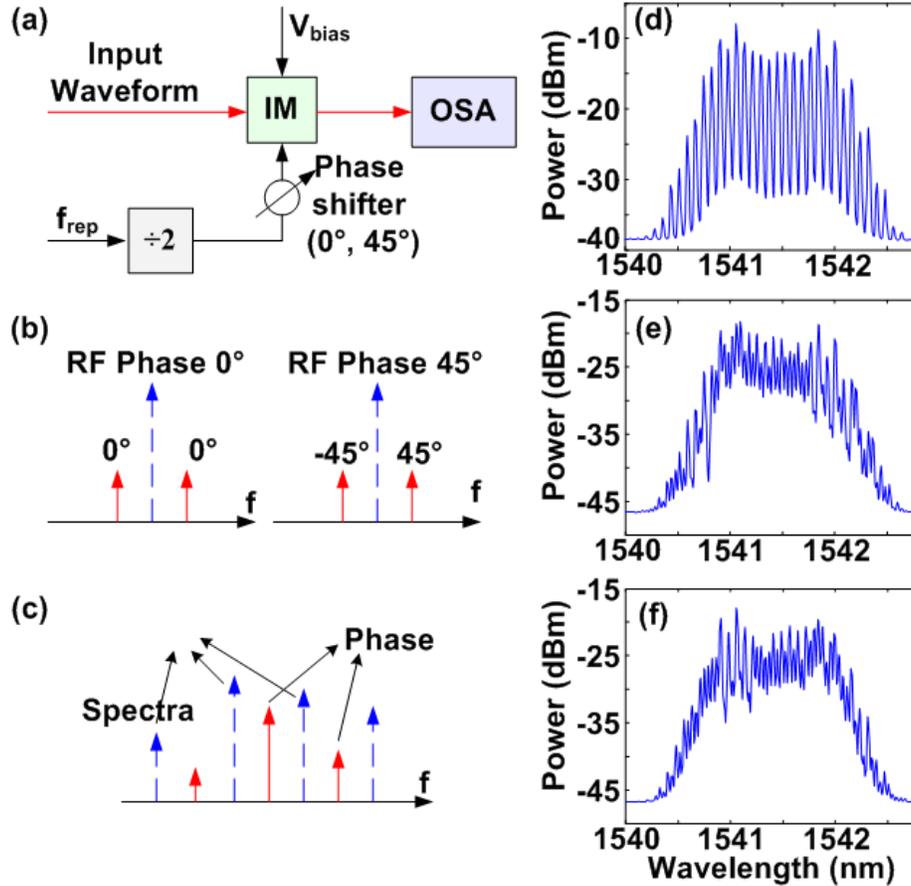

Fig.1 (a) Experimental setup, IM – Intensity modulator, (b) Schematic indicating phase relations between 1$^{st}$ order sidebands as the phase of the modulating RF sinusoid is changed, (c) Schematic indicating the situation when a frequency comb undergoes modulation. The amplitudes at the carrier positions are used to obtain the amplitude spectra while the interference between sidebands is used to obtain the phase information, (d) Measured spectra from the OSA with no modulation, (e), (f) Measured OSA spectra with modulation on for two settings of the RF phase shifter spaced 45° apart.

The expressions for the interference between the sidebands of two adjacent lines (say $n$ and $n+1$) are

$$I_{inphase} = C[|a_n|^2 + |a_{n+1}|^2 + 2|a_n||a_{n+1}|\cos(\psi_n - \psi_{n+1})] \qquad (1)$$

$$I_{quadrature} = C[|a_n|^2 + |a_{n+1}|^2 + 2|a_n||a_{n+1}|\sin(\psi_n - \psi_{n+1})] \qquad (2)$$

Where $|a_i|$ and $\psi_i$ are the spectral amplitude and phase of the input comb lines and $C$ is a modulation parameter defined as the ratio between the sideband and the carrier powers for a single frequency input. By using the sideband power for the two RF phase shifts and the power of the two associated carrier components, both the constant $C$ and the spectral phase difference between adjacent lines may be obtained by solving equations (1-2). Here we make the simple assumption that $C$ remains relatively unchanged in the span of two comb lines. In principle $C$ may vary slowly across the optical spectrum without degrading the phase measurement. Once the spectral phase difference between adjacent lines is obtained, it can then be summed as in conventional SSI to recover the spectral phase. In the above equations we neglected a constant offset phase in both equations (we assumed them to be *cosine* and *sine* respectively). However this corresponds only to a linear term in the retrieved phase, which we ignore since it corresponds to a simple pulse delay. A necessary requirement for our scheme to work well is to have a low enough RF drive voltage such that the $2^{nd}$ order sidebands (which occur at the same position as the next carrier) do not affect the measurement ($V/V_\pi < 0.05$ to get the $2^{nd}$ order power < 40dB compared to the carrier from which it is formed). However, with such a low drive voltage there is a possibility that spectrometer crosstalk from the carrier can adversely affect the $1^{st}$ order sidebands. For example, the OSA we use (ANDO, 1.25 GHz specified spectral resolution) has a crosstalk of -25dB at 5 GHz offset, which is the position of the first order sidebands in our case. This will be a significant concern if we use a phase modulator. But with the IM, the bias comes in handy to ensure that the carrier is suppressed enough to avoid overwhelming the sidebands with incoherent crosstalk. We adjusted the IM settings to get >40dB power ratio between the carrier and the 2nd order side band and a carrier to first order sideband power ratio of ~10dB. These settings can be easily verified if necessary by using a single CW input to the modulator. In contrast, in reference [18] the IM was set for full carrier suppression, which prevents access to the amplitude spectrum during the same acquisition. Because of this, four measurements were needed with different phase shift settings, rather than two in our case. From a practical perspective, we have observed that the amplitude spectrum of some comb sources can drift relatively quickly. Measurement accuracy is significantly improved when amplitude and phase related information is acquired in the same acquisition.

Before we go into the results with our scheme, we would like to point out some other related techniques. Electro-optic spectral shearing interferometry [12] is a linear variant of SSI which uses a phase modulator to achieve spectral shearing. Though cosmetically similar, this technique is fundamentally different from ours. There, a pulse pair is created and sent through a phase modulator with the relative delay of the pulse pair chosen so that spectral shearing occurs on opposing linear sections of the sinusoid. This technique is primarily aimed at short pulses and is subject to the same limitations with respect to the 100% duty factor waveforms common in OAWG as conventional SSI. Another technique more closely related to what we use is discussed in [19], where instead of using different settings on the phase shifter, the modulation frequency is slightly detuned from half the repetition rate, and lock-in detection is used to obtain the coherent interference. However, this technique requires isolating individual pairs of comb lines using a monochromator; full phase information is obtained by sweeping across the comb spectrum. Hence, this approach becomes time consuming as the bandwidths of the waveforms become larger or the power in each comb line become smaller. As mentioned previously, since in our approach we need exactly two measurements (which can further be improved to a single measurement using a dual-channel spectrometer), we believe our technique provides a relatively simpler and quicker measurement technique.

## 3. Results

Figure 1(d) shows a representative spectrum from a comb source generated by strong phase and intensity modulation of a CW laser [5]. Since the phase modulation dominates, in the time domain this comb still has a relatively flat, wide temporal envelope. This corresponds to abrupt line to line phase variations. Figs 1(e) and 1(f) show representative spectra obtained with the two RF phase shifter settings from which the spectral amplitude and phase can be obtained. Fig 2(a) shows the retrieved spectral phase indicating a strong line to line phase variation. By programming the retrieved phase onto a pulse-shaper [1], we compress the comb signal into a train of bandwidth-limited pulses. Figure 2(b) shows the intensity autocorrelations of different cases superimposed on each other. The solid line (green) corresponds to the uncorrected comb having a wide envelope with close to 100% duty factor, which is a hallmark of OAWG waveforms. Phase retrieval followed by correction with a pulse shaper was done at two different average power levels (measured at the input to the measurement apparatus) (~3mW (dotted, blue) and 100nW (dashed, red)). Though different powers were used for the measurement, all the three autocorrelations were taken at the same input power to the autocorrelator and have been normalized relative to each other. In both cases involving correction, the autocorrelation clearly collapses into a pulse, which validates the measurement results. We further verified the measured autocorrelations with the simulated autocorrelation taking the spectra into account and assuming flat spectral phase and we observed that, in the 3mW case, the measured and simulated autocorrelations match very well. Regarding powers, we note that at 100nW, the energy per pulse is 10aJ, indicating very low power operation. The measurement time is largely limited at this point by the sweep speed of the monochromator based OSA which is of the order of a few seconds. However, by moving to a spectrometer with a detector array as in [13], this can be significantly improved (we have shown acquisitions as fast as ~1.4 microseconds in [13]).

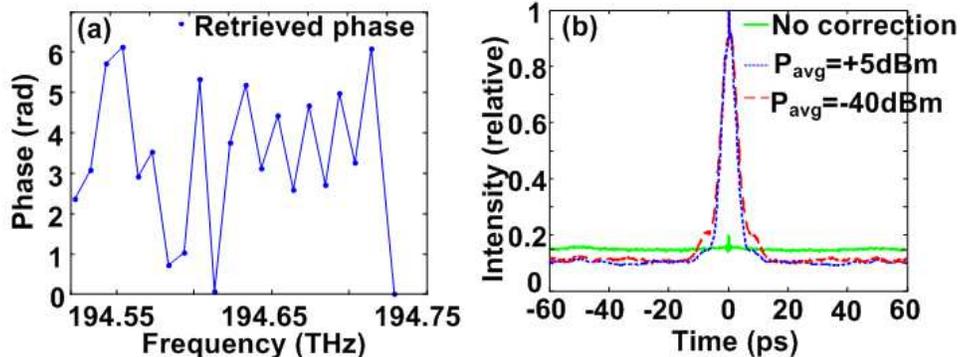

Fig.2 (a) Retrieved spectral phase for the comb source, (b) Measured intensity autocorrelations of the comb source with no spectral phase correction (solid line) and for phase correction based on measurements performed at an average power of 0dBm (dotted line, blue), -40dBm (dashed line, red). All the autocorrelations were performed with the same average input power into the autocorrelator. The simulated intensity autocorrelation (not plotted) taking the measured power spectra and assuming flat spectral phase matches very closely with the 0dBm trace.

Figure 3 shows experimental results using a different comb source, namely an optical frequency comb generator (OFCG) [2]. The OFCG generates multiple sidebands of an input CW source by driving a phase modulator in a Fabry-Perot cavity. Pulses are created whenever the Fabry-Perot cavity is resonant with the input laser. A properly biased cavity will produce two evenly spaced but frequency shifted pulses per RF period when driven with a sinusoidal source [20]. When biased incorrectly, the bandwidth of the OFCG comb decreases, and the timing of the optical pulses changes. Fig 3(a) and 3(b) show the measured spectrum and spectral phase for a properly biased OFCG, producing ~50-ps pulse to pulse spacing (one half of the period for a 10-GHz comb). The positive slope of one half of the spectrum and the negative slope of the other half clearly shows not only the intensity information of having two pulses per roundtrip but the complete information showing that they occupy different parts of the source spectrum. Fig 3(c) shows an interesting aspect of the retrieved phase. The phase shown is the part circled in fig 3(b), but now the linear part corresponding to the delay is removed. We see that the phase agrees very well with a quadratic fit (standard deviation of error $< 0.02\pi$) corresponding to ~35m of standard SMF. This is due to the fiber link between the OFCG source and the measurement apparatus. This data further indicates the precision of our experimental setup. Also, due to the quadratic chirp acting on the two pulses which are separated in frequency, the spacing between them is expected to slightly reduce as they propagate over fiber [20].

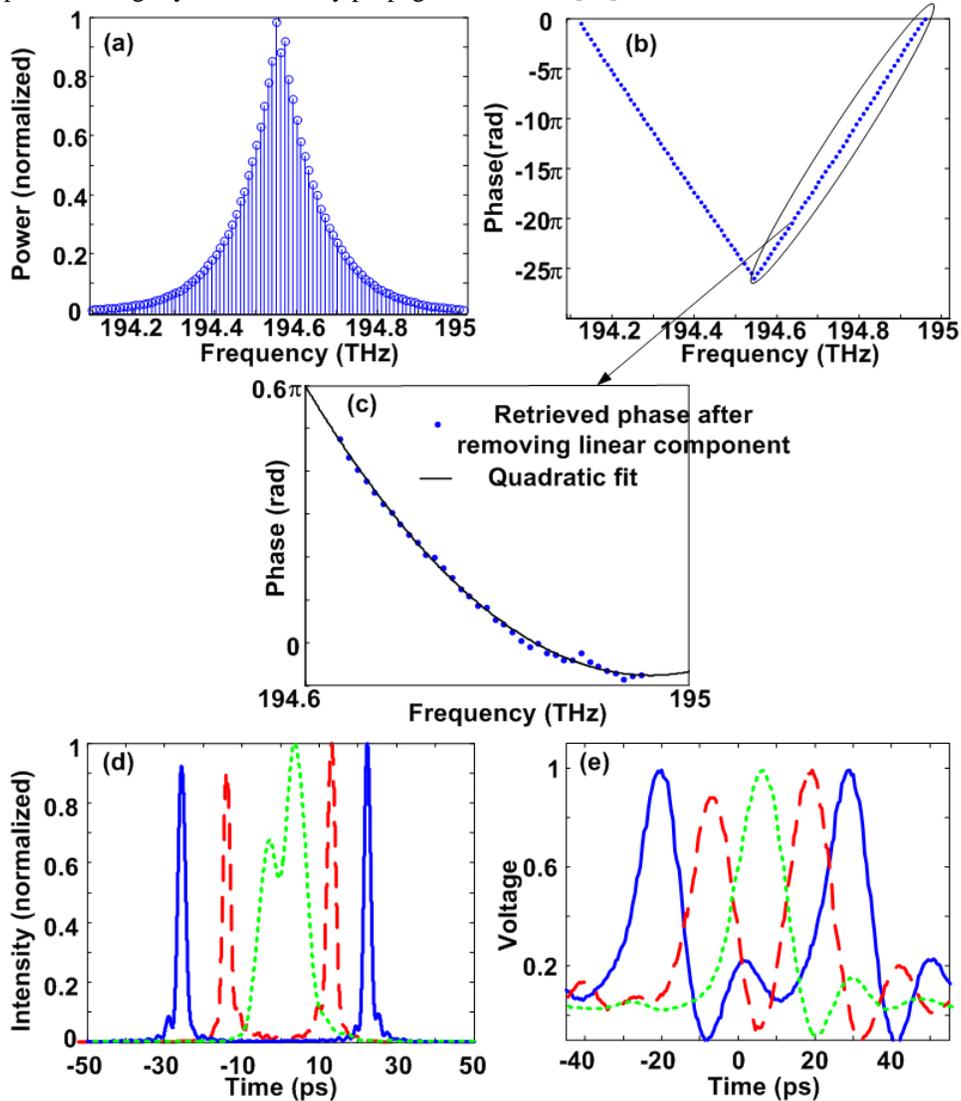

Fig.3 (a), (b) Retrieved spectra and spectral phase for the OFCG at an optimal setting with the pulses spaced at half the repetition period, (c) Plot showing the spectral phase circled in 3(b) corresponding to one of the pulses after the linear component corresponding to delay is subtracted and a quadratic fit to it (d) Recovered time domain intensities from the measured spectra and spectral phase, (solid line, blue) corresponding to spectra and phase shown in 3(a) and 3(b), (dashed line, red), (dotted line, green) with changing bias conditions from optimality, (e) Corresponding scope traces obtained with a 50GHz photodiode and sampling scope.

Simulations showed us that this number is ~2ps. The retrieved time domain waveform from the amplitude and phase information (by a Fourier transform relation) is shown in fig 3(d) (solid, blue). On closer inspection, we saw that the pulse spacing was smaller than 50ps by around 2ps indicating the chirp effect discussed above. For comparison we show the measured waveform with a 60GHz photodiode and 50 GHz sampling oscilloscope in fig 3(e) (solid blue). We see a clear match between the two, but owing to the much larger measurement bandwidth (>1 THz) in the optical measurement case, the true pulse shapes are seen, which include the actual temporal width and any temporal dispersion effects. We can modify the output waveforms by changing the bias voltage of the RF drive signal to the OFCG, shown in figs 3(d) and 3(e) ((dashed, red), (dotted, green)). The pulse positions revealed by our optical method and electrical detection are again in good agreement, although the optical measurement provides much finer temporal resolution. Figs 3(d) and 3(e) (dotted, green) are particularly interesting because in this setting, the bandwidth of the OFCG is smaller (corresponding to a wider pulse), and the pulses begin to overlap; these effects cannot be discerned with the scope trace.

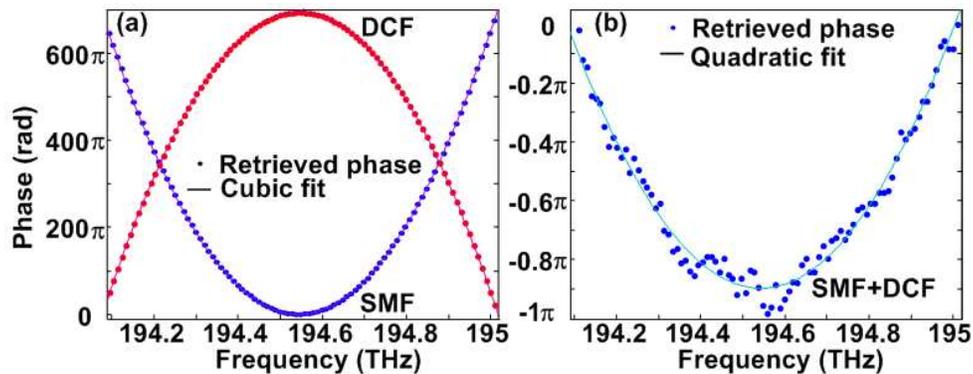

Fig.4 (a) Retrieved spectral phase with cubic fits for a SMF link of ~25km and a DCF module matched to it, (b) Retrieved spectral phase and quadratic fit for the dispersion compensated (SMF+DCF) link

Our next experiment was to measure spectral phases as the OFCG frequency comb undergoes dispersive propagation over long lengths of fiber. To simulate the situation where the two ends of the fiber are at two different locations, we derived the repetition rate from the signal itself, by sending a small fraction to a 20 GHz photodiode and band pass filtering for clock recovery. Figure 4(a) shows the retrieved spectral phase and the cubic fits for an ~25 km spool of standard single mode fiber (SMF) fiber and a matched dispersion compensating fiber (DCF) module (vendor – OFS-Fitel). The standard deviation of the error between the fit and retrieved phase is small (<0.05 $\pi$), and the obtained dispersion and dispersion slope parameters from the fit (SMF: 393.8ps/nm, 1.5ps/nm^2) and (DCF: -392.1ps/nm, -1.6ps/nm^2) agree well with vendor specifications and our previous measurements [21]. As a further check we measured the residual phase for the dispersion compensated link in which the SMF and DCF were connected in series (fig 4(b)). The obtained residual dispersion of 1.2ps/nm is close to the expected value of 1.7ps/nm (obtained by taking the difference between the dispersions of SMF and DCF measured individually). This provides a further strong validation of our scheme and demonstrates both high measurement precision and dynamic range.

In summary, in this work we reported experiments on pulse compression and characterization of two different "novel" frequency comb sources as well as measurements of fiber dispersion by means of an easy-to-use and linear adaptation of zero-delay spectral shearing interferometry. We observed good performance down to very low power levels (100nW average, 10aJ/pulse). This technique may be generalized to low repetition rate sources, such as passively mode-locked femtosecond fiber lasers, either by using phase locked oscillators to generate spectral shears which are multiples of the repetition rate or by using high resolution OSAs [22] to achieve line-by-line characterization of these waveforms. As we move towards lower repetition combs or wider bandwidths or both, integrated noise may be expected to increasingly affect the phase retrieval process, as in conventional SSI [11]; further investigation is necessary in order to quantify the limits for scaling of this measurement technique.

**Acknowledgement**
We would like to thank the NSF under grant ECCS-0601692